**Title**: Logistic approximations used to describe new outbreaks in the 2020 COVID-19 pandemic

**Authors**: Apiano F. Morais

**Affiliations**: Physics Department, Universidade Regional do Cariri, Juazeiro do Norte, Brazil

**Contact email**: apiano.morais@urca.br

**Abstract**

In this investigation I used the Logistic Model to fit the COVID-19 pandemic data for some countries. The data modeled is the death numbers per day in China, Iran, Italy, South Korea, Spain and United States. Considering the current growth rate of the pandemic, it was possible to show that the death toll should be between 3,277-3,327 deaths in China, 2,035-2,107 in Iran, 120-134 in South Korea, 11,227-12,793 in Italy and 6,217-7,405 in Spain. Also, with this toy model it was possible to show a clear emergence of a new outbreak within the same country (Iran, China and the United States). The growth rate of deaths found for South Korea was the lowest among the countries studied ($0.14701\pm0.00923$) and for China ($0.16667\pm0.00284$). Italy ($0.22594\pm0.00599$) and Spain ($0.31213\pm0.02337$) had the highest rates and in the second wave in Iran ($0.37893\pm0.02712$).

# Introduction

The disease called COVID-19 was initially identified in the city of Wuhan, in the province of Hubei in China, in November 2019 [1]. The infection quickly spread to Asian countries and reached all other continents in early February [2]. The form of contagion is mainly through respiratory and blood contact with the coronavirus. Studies have shown that the virus can survive for long hours on different surfaces. This makes the virus easily transported on airplanes and boats by uninfected goods or infected people [3]. The disease is more lethal for patients with compromised health conditions and among the elderly [4].

Contrary to the World Health Organization (WHO), which emphasized the need to test all suspected cases, some countries have chosen not to do this. Also, those countries authorities ordered to their citizens, who are not in the risk groups, do not go for medical help. This leads to a large number of sub-notifications of the disease. Thus, the number of infected in each country is an unreliable data, since governments apply different methodologies in testing the disease in the population. One of the reasons why the mortality rate for COVID-19 is different around the world is precisely the under-reporting. However, when the patient dies, in most cases, health authorities perform COVID-19 tests on the deceased. These data are not only more reliable in terms of the number of deceased, but also in relation to the date of death.

## Model

The data used in this investigation, I obtained through access to the data bank of the World Health Organization (WHO) on the submission date of this paper [5]. Once the epidemic starts, governments take steps to reduce the rate of infection. I conjectured in this article that the death rate is not greatly altered by government actions.

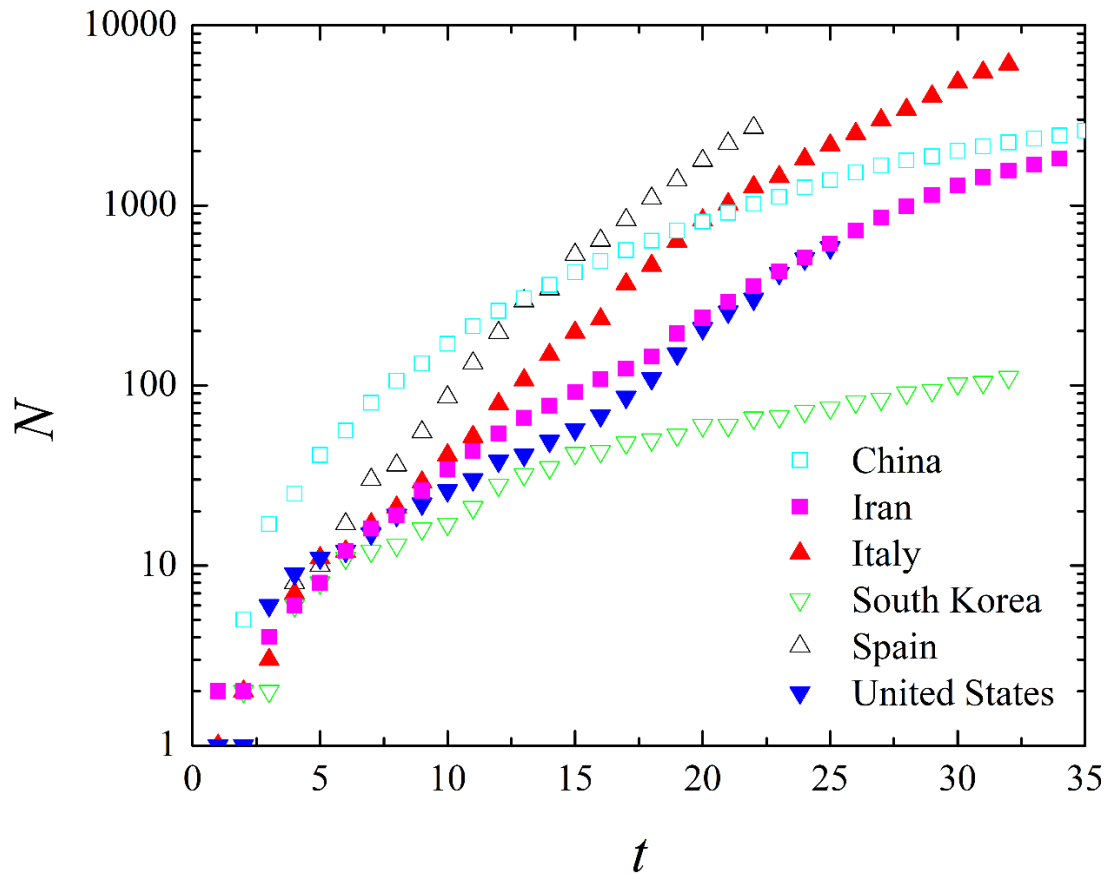

*Figure 1 - Cumulative total deaths, N, versus days since the first demise, t, for several countries.*

Physicists, biologists and mathematicians have a long tradition of studying the behavior of population growth and their interactions [6]–[9]. The beginning of the epidemic is the most difficult phase to model, as the models generally use the consideration of a continuous population field. Furthermore, the reliability of data for small numbers does not guarantee a good agreement between data and models. Obviously, different diseases result in different outbreaks according to the health conditions of the population, type of contagion, climate and so on.

Perhaps the most important factor, which is often overlooked in epidemic studies, is the topology of the population interaction network. Recent models for infections and vaccination require the actions of individuals to be considered, as well as the topology of society [10]. Clearly, no model based on stationary agents can accurately describe reality. A mix between dynamic and stationary models has shown that zero order models are a good approximation when there are already large numbers of infected (dead) people [11]–[13].

Several zero-order population growth models for epidemics have been proposed [14]–[16] since the Verhulst logistic model [17]. Recently, Castorina and colleagues [18] have applied similar methods to data for infections. And they have basically the same characteristics to determine the population number $N$ (in this case, the number of deceased): a presumed equilibrium $K$, the growth rate $r$ and the initial number $N_0 = N(t = 0)$ of infected (dead). $K$ in this model represents the saturated population of deceased, i.e. the limit of individuals in the risk group. $r$ here is expressed in 1/day units.

The rate at which the deceased population grows depends on the country's healing capacity and the size of the risk group:

$$\frac{dN}{dt} = rN\left(1 - \frac{N}{K}\right).$$

Equation 1

This equation solution is a three-parametric function:

$$N(t) = \frac{K}{1 + Ce^{-kt}},$$

Equation 2

where $C \equiv (K - N_0)/N_0$.

In this paper I used equation 2 to fit the WHO reported data for some countries in order to obtain $K$ and $r$.

## Results and Discussion

I used the OriginPro 7.5 software for data processing and curve fitting. The curves were adjusted using a simplex iterative method with at least 1000 iterations with convergence criteria for the parameters as low as $10^{-9}$.

In the case of Italy, the death toll is starting to drop, but has not yet reached equilibrium. The projection of balance, according to the model, is frightening ($K = 12{,}010.76 \pm 783.39$ deceased and $r = 0.22594 \pm 0.00599$).

The epidemic in Spain has grown at a higher rate than in Italy, compared to the early days. However, the rate of growth in the number of deceased has decreased. The balance in Spain, according to the model at the current pace, is expected to reach $K = 6{,}811.95 \pm 593.711$ deaths ($r = 0.31213 \pm 0.02337$).

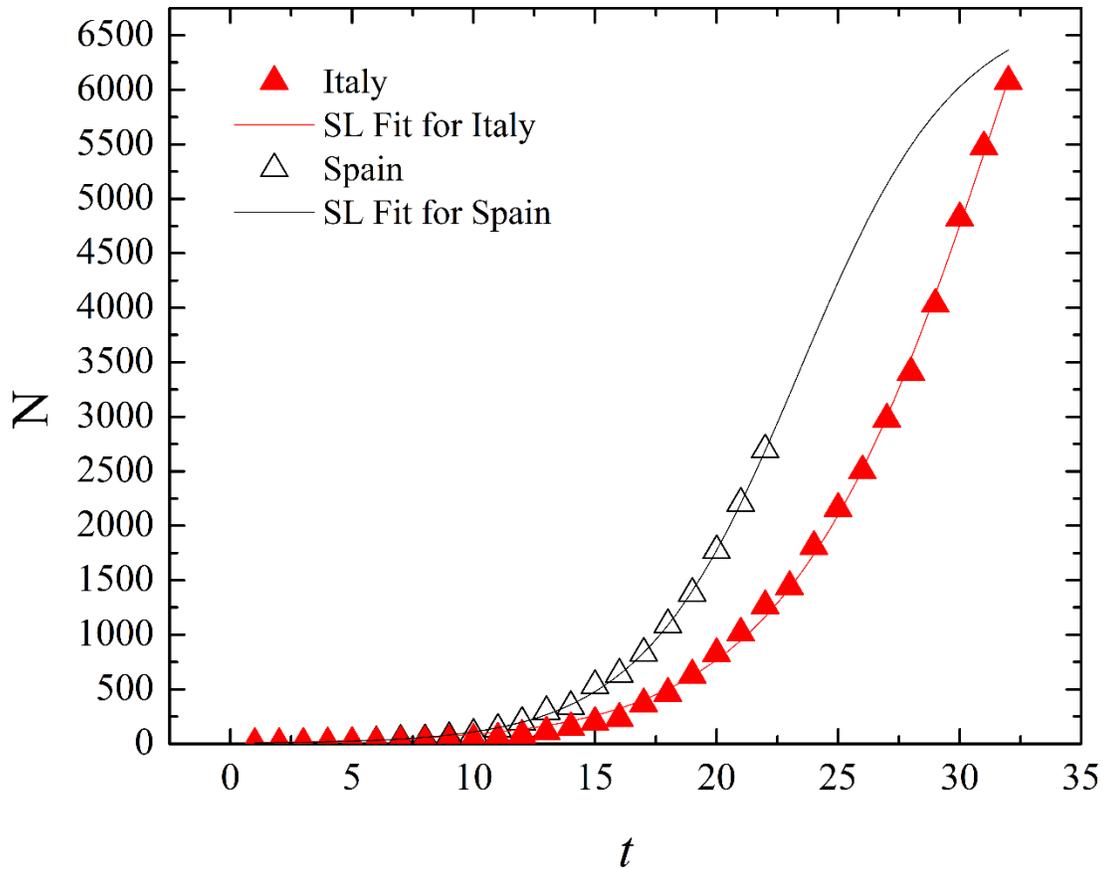

*Figure 2- Cumulative total deaths, N, versus days since the first demise, t, for Italy and Spain. Although visually the growth rate of deaths in Spain appears to be higher than in Italy, the model's curve adjustments show that the equilibrium in Spain is lower than in Italy.*

Since the infections is a geographic issue, as new sites are infected, it is possible that new epidemic center takes place. Thus, countries with controlled epidemic centers may suffer from the outbreak in a new region within the country. Clearly, the larger the country, the more susceptible the creation of new outbreaks of the epidemic.

The first deaths from the disease came in Iran, Italy and South Korea at about the same time. However, it is possible to notice in figure 1 that the evolution of the victims followed quite different dynamics. Korean authorities have decided to test all suspected cases and individuals at risky occupations. This way quarantine has not been applied in South Korea. Korean government tested more than 270,000 people so far [19]. I adjusted the curves of the Korean data resulting in the following parameters: $K = 127.44 \pm 6.99637$ deceased and $r = 0.14701 \pm 0.00923$. The epidemic appears to be under control in China, South Korea, Japan and Singapore.

Since Korean data for the lethality of the disease are more reliable, I used the lethality rate of 5.27\% as the value for the age group 65 and older [20]. the population of Lombardy, the center of the disease in Italy, is around 10 million people. Of this total, 23.1% are over the age of 64, which puts 2 million people there in the group at risk for the disease [21].This means that 105 thousand people would be deceased without treatment in Italy, due to the extrapolation of the model.

Italy (301,338 km²) and South Korea (100,210 km²) are populous countries, but small compared to Iran (1,648 million km²). The likelihood of having more than one epidemic center in Iran is higher and the model shows. In this case, each epidemic center has its own growth curve and the country's curve is the sum of the two curves. At this point, the equation is modified to accommodate a time difference, $\tau$, between the two centers:

$$N(t) = \frac{K}{1 + Ce^{-k(t-\tau)}},$$

Equation 3

The second outbreak in Iran is quite remarkable from the data presented in Figure 3. I obtained the parameters for the first outbreak in Iran as $K = 857.97 \pm 598.19$ deceased and $r = 0.21497 \pm 0.01974$. For the second wave, the parameters obtained are $K = 1214.93251 \pm 36.42229$ deceased and $r = 0.37893 \pm 0.02712$ with $\tau = 22.48 \pm 29.35$. The growth curve of deaths in the United States also suggests the presence of more than one epidemic center as reported with most deaths in states of New York, Washington D.C. and California. This makes modeling through this toy model not possible yet.

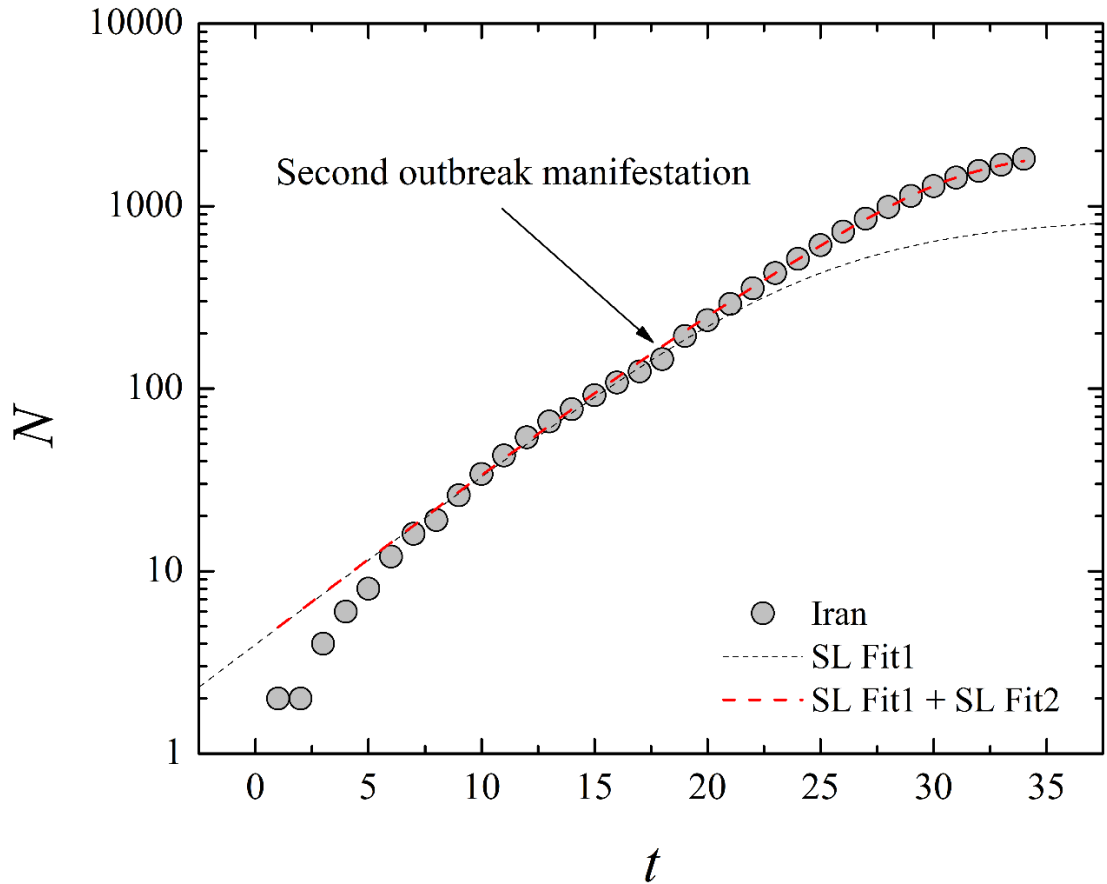

*Figure 3- Cumulative total deaths, N, versus days since the first demise, t, for Iran. The data shows a clear manifestation of a second outbreak around the 19th day.*

There appears to be a change in behavior on the curve on the thirty-ninth day after the deaths began (see Figure 4). I suggest that this could be a new outbreak outside Wuhan. I characterized the main outbreak of the disease in China with the balance at $K = 3170.76 \pm 40.10$ deceased and a rate of $r = 0.16667 \pm 0.00284$. I characterized the second outbreak of the disease with the following parameters $K = 103.795 \pm 13.62$ deceased and $r = 0.38478 \pm 0.6480$. I estimate, by the model, that if there are no new epidemic centers, the number of deaths in China will be between 3,277 and 3,327. The adjustment for the start day of the second outbreak presents a very large error due to the small number of data.

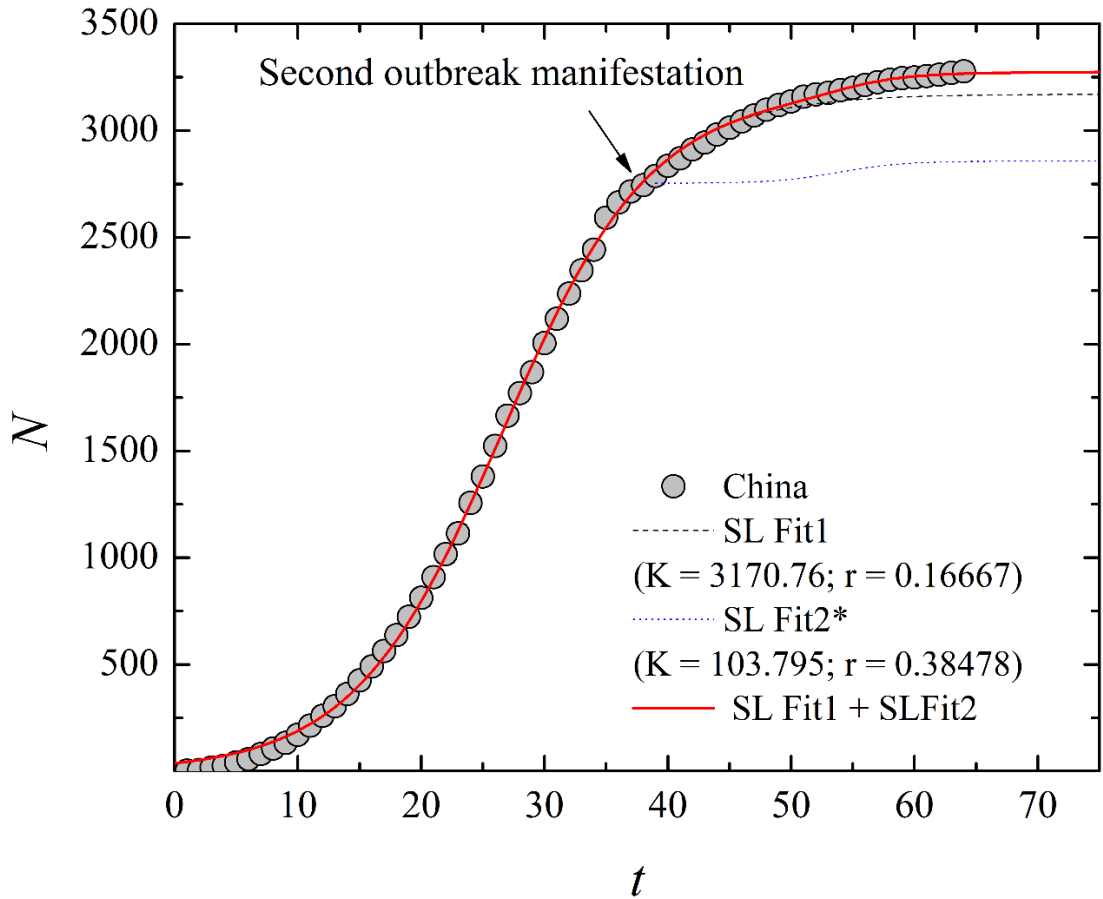

*Figure 4- Cumulative total deaths, N, versus days since the first demise, $r$, for China. The data are better adjusted when we add two logistic curves. This shows that there may have been a second outbreak. The adjustment of the second outbreak in the graph is shown plus the data value for the thirty-ninth day for aesthetic effects*

If the same conditions are maintained, the total number of deaths per country can be seen in table 1 as the Presumed Equilibrium $K$.

*Table 1- Presumed Equilibrium K and growth rate $r$ for the studied countries.*

| Country | Presumed Equilibrium (deaths) | $r$ (1/day) |
|---|---|---|
| **China** | 3,277-3,327 | not defined |
| **Iran** | 2,035-2,107 | not defined |
| **Italy** | 11,227-12,793 | 0.22594±0.00599 |
| **South Korea** | 120-134 | 0.14701±0.00923 |
| **Spain** | 6,217-7,405 | 0.31213±0.02337 |

# Conclusions

Clearly, this is a toy model for describing the increase in the population of dead in an epidemic. Conjectures such as a fixed growth rate of mortality, $r$, can only be assumed if health authorities do not act or a very rapid evolution of the disease, which may be the case with COVID-19. However, several models proposed for the growth of deaths follow an equation similar to that of Verhulst.

It was possible to adjust various data with the logistic model. I showed that for countries of great territorial extension, several epidemic outbreaks can occur. This makes the georeferenced notification of fundamental importance in order to be able to define the equilibrium population and the growth rate of the epidemic in that region. Part of Europe, the Americas, Oceania and Africa are still facing the start of the pandemic and the data suggest that the measures adopted by China and South Korea should be strongly replicated when possible in order to slow the rate of growth of the deceased.